\pdfoutput=1
\documentclass[12pt,a4paper]{article}

\usepackage{booktabs} 
\usepackage{multirow}
\usepackage{bm}
\usepackage{lscape}

\usepackage{ifthen} 
\newboolean{pdflatex}
\setboolean{pdflatex}{true} 

\newboolean{articletitles}
\setboolean{articletitles}{true} 

\newboolean{uprightparticles}
\setboolean{uprightparticles}{false} 

\def\paperauthors{LHCb collaboration} 
\def\paperasciititle{First observation of the Lambdab0 to D+ D- Lambda0 decay} 
\def\papertitle{First observation of 
the $\Lb\to\Dp\Dm\Lz$ decay} 
\def\paperkeywords{{High Energy Physics}, {LHCb}} 
\def\papercopyright{\the\year\ CERN for the benefit of the LHCb collaboration}
\def\paperlicence{CC BY 4.0 licence}
\def\paperlicenceurl{https://creativecommons.org/licenses/by/4.0/}

\usepackage[top=1in, bottom=1.25in, left=1in, right=1in]{geometry}

\usepackage{microtype}
\usepackage{lineno}  
\usepackage{xspace} 
\usepackage{caption} 

\usepackage{graphicx}  
\usepackage{color}
\usepackage{colortbl}
\graphicspath{{./figs/}} 

\usepackage{amsmath} 
\usepackage{amssymb}
\usepackage{amsfonts}
\usepackage{upgreek} 

\newcommand*\patchAmsMathEnvironmentForLineno[1]{%
\expandafter\let\csname old#1\expandafter\endcsname\csname #1\endcsname
\expandafter\let\csname oldend#1\expandafter\endcsname\csname
end#1\endcsname
 \renewenvironment{#1}%
   {\linenomath\csname old#1\endcsname}%
   {\csname oldend#1\endcsname\endlinenomath}%
}
\newcommand*\patchBothAmsMathEnvironmentsForLineno[1]{%
  \patchAmsMathEnvironmentForLineno{#1}%
  \patchAmsMathEnvironmentForLineno{#1*}%
}
\AtBeginDocument{%
\patchBothAmsMathEnvironmentsForLineno{equation}%
\patchBothAmsMathEnvironmentsForLineno{align}%
\patchBothAmsMathEnvironmentsForLineno{flalign}%
\patchBothAmsMathEnvironmentsForLineno{alignat}%
\patchBothAmsMathEnvironmentsForLineno{gather}%
\patchBothAmsMathEnvironmentsForLineno{multline}%
\patchBothAmsMathEnvironmentsForLineno{eqnarray}%
}

\usepackage{hyperxmp}

\usepackage[pdftex,
            pdfauthor={\paperauthors},
            pdftitle={\paperasciititle},
            pdfkeywords={\paperkeywords},
            pdfcopyright={Copyright (C) \papercopyright},
            pdflicenseurl={\paperlicenceurl}]{hyperref}

\usepackage[colorinlistoftodos,textsize=scriptsize]{todonotes}

\usepackage[bottom,flushmargin,hang,multiple]{footmisc}

\usepackage[all]{hypcap} 

\usepackage{xspace} 
\usepackage{upgreek}

\def\lhcb   {\mbox{LHCb}\xspace}

\def\MagUp {\mbox{\em Mag\kern -0.05em Up}\xspace}

\ifthenelse{\boolean{uprightparticles}}%
{

 \def\Ppi         {\ensuremath{\uppi}\xspace}

 \def\PDelta      {\ensuremath{\Delta}\xspace}                 
 \def\PXi         {\ensuremath{\Xi}\xspace}                 
 \def\PLambda     {\ensuremath{\Lambda}\xspace}                 
 \def\PSigma      {\ensuremath{\Sigma}\xspace}                 
 \def\POmega      {\ensuremath{\Omega}\xspace}                 
 \def\PUpsilon    {\ensuremath{\Upsilon}\xspace}
 \let\oldPi\Pi
 \def\PPi         {\ensuremath{\oldPi}\xspace}

 \def\PB      {\ensuremath{\mathrm{B}}\xspace}                 
                  
 \def\PD      {\ensuremath{\mathrm{D}}\xspace}

 \def\PK      {\ensuremath{\mathrm{K}}\xspace}

 \def\Pb      {\ensuremath{\mathrm{b}}\xspace}                 
 \def\Pc      {\ensuremath{\mathrm{c}}\xspace}                 
 \def\Pd      {\ensuremath{\mathrm{d}}\xspace}

 \def\Pi      {\ensuremath{\mathrm{i}}\xspace}

 \def\Pp      {\ensuremath{\mathrm{p}}\xspace}

 \def\Ps      {\ensuremath{\mathrm{s}}\xspace}                 
                  
 \def\Pu      {\ensuremath{\mathrm{u}}\xspace}

 \def\thebaroffset{0.0em}
}
{

 \def\Ppi         {\ensuremath{\pi}\xspace}

 \mathchardef\PDelta="7101
 \mathchardef\PXi="7104
 \mathchardef\PLambda="7103
 \mathchardef\PSigma="7106
 \mathchardef\POmega="710A
 \mathchardef\PUpsilon="7107
 \mathchardef\PPi="7105
                  
 \def\PB      {\ensuremath{B}\xspace}                 
                  
 \def\PD      {\ensuremath{D}\xspace}

 \def\PK      {\ensuremath{K}\xspace}

 \def\Pb      {\ensuremath{b}\xspace}                 
 \def\Pc      {\ensuremath{c}\xspace}                 
 \def\Pd      {\ensuremath{d}\xspace}

 \def\Pi      {\ensuremath{i}\xspace}

 \def\Pp      {\ensuremath{p}\xspace}

 \def\Ps      {\ensuremath{s}\xspace}                 
                  
 \def\Pu      {\ensuremath{u}\xspace}

 \def\thebaroffset{0.18em}
}
\newcommand{\offsetoverline}[2][\thebaroffset]{\kern #1\overline{\kern -#1 #2}}%

\makeatletter
\ifcase \@ptsize \relax
  \newcommand{\miniscule}{\@setfontsize\miniscule{4}{5}}
\or
  \newcommand{\miniscule}{\@setfontsize\miniscule{5}{6}}
\or
  \newcommand{\miniscule}{\@setfontsize\miniscule{5}{6}}
\fi
\makeatother

\DeclareRobustCommand{\optbar}[1]{\shortstack{{\miniscule (\rule[.5ex]{1.25em}{.18mm})}
  \\ [-.7ex] $#1$}}

\def\uquark    {{\ensuremath{\Pu}}\xspace}

\def\dquark    {{\ensuremath{\Pd}}\xspace}

\def\squark    {{\ensuremath{\Ps}}\xspace}

\def\cquark    {{\ensuremath{\Pc}}\xspace}
\def\cquarkbar {{\ensuremath{\overline \cquark}}\xspace}
\def\ccbar     {{\ensuremath{\cquark\cquarkbar}}\xspace}
\def\bquark    {{\ensuremath{\Pb}}\xspace}

\def\pion   {{\ensuremath{\Ppi}}\xspace}
\def\piz    {{\ensuremath{\pion^0}}\xspace}
\def\pip    {{\ensuremath{\pion^+}}\xspace}
\def\pim    {{\ensuremath{\pion^-}}\xspace}

\def\kaon    {{\ensuremath{\PK}}\xspace}

\def\KorKbar {\kern \thebaroffset\optbar{\kern -\thebaroffset \PK}{}\xspace}
\def\Kz      {{\ensuremath{\kaon^0}}\xspace}

\def\Kp      {{\ensuremath{\kaon^+}}\xspace}
\def\Km      {{\ensuremath{\kaon^-}}\xspace}

\def\KS      {{\ensuremath{\kaon^0_{\mathrm{S}}}}\xspace}
\def\Vzero   {{\ensuremath{V^0}}\xspace}
\def\KL      {{\ensuremath{\kaon^0_{\mathrm{L}}}}\xspace}

\def\Dbar    {{\ensuremath{\offsetoverline{\PD}}}\xspace}
\def\D       {{\ensuremath{\PD}}\xspace}

\def\DorDbar {\kern \thebaroffset\optbar{\kern -\thebaroffset \PD}\xspace}
\def\Dz      {{\ensuremath{\D^0}}\xspace}
\def\Dzb     {{\ensuremath{\Dbar{}^0}}\xspace}
\def\Dp      {{\ensuremath{\D^+}}\xspace}
\def\Dm      {{\ensuremath{\D^-}}\xspace}
\def\Dpm     {{\ensuremath{\D^\pm}}\xspace}
\def\Dmp     {{\ensuremath{\D^\mp}}\xspace}
\def\DpDm    {\ensuremath{\Dp {\kern -0.16em \Dm}}\xspace}

\def\Dstarp  {{\ensuremath{\D^{*+}}}\xspace}
\def\Dstarm  {{\ensuremath{\D^{*-}}}\xspace}

\def\B       {{\ensuremath{\PB}}\xspace}

\def\BorBbar {\kern \thebaroffset\optbar{\kern -\thebaroffset \PB}\xspace}
\def\Bz      {{\ensuremath{\B^0}}\xspace}

\def\Bd      {{\ensuremath{\B^0}}\xspace}

\def\BdorBdbar {\kern \thebaroffset\optbar{\kern -\thebaroffset \Bd}\xspace}

\def\Bs      {{\ensuremath{\B^0_\squark}}\xspace}

\def\BsorBsbar {\kern \thebaroffset\optbar{\kern -\thebaroffset \Bs}\xspace}

\def\Y#1S{\ensuremath{\PUpsilon{(#1S)}}\xspace}

\def\proton      {{\ensuremath{\Pp}}\xspace}

\def\Lz          {{\ensuremath{\PLambda}}\xspace}

\def\LorLbar     {\kern \thebaroffset\optbar{\kern -\thebaroffset \PLambda}\xspace}

\def\Sigmares    {{\ensuremath{\PSigma}}\xspace}

\def\Xires       {{\ensuremath{\PXi}}\xspace}

\def\Xic         {{\ensuremath{\Xires_\cquark}}\xspace}

\def\Lb           {{\ensuremath{\Lz^0_\bquark}}\xspace}

\def\BF         {{\ensuremath{\mathcal{B}}}\xspace}
\def\BR         {\BF}

\def\to                 {\ensuremath{\rightarrow}\xspace}

\def\AT#1     {\ensuremath{A_{\mathrm{T}}^{#1}}\xspace}           

\def\C#1      {\ensuremath{\mathcal{C}_{#1}}\xspace}                       
\def\Cp#1     {\ensuremath{\mathcal{C}_{#1}^{'}}\xspace}                    
\def\Ceff#1   {\ensuremath{\mathcal{C}_{#1}^{\mathrm{(eff)}}}\xspace}        
\def\Cpeff#1  {\ensuremath{\mathcal{C}_{#1}^{'\mathrm{(eff)}}}\xspace}       
\def\Ope#1    {\ensuremath{\mathcal{O}_{#1}}\xspace}                       
\def\Opep#1   {\ensuremath{\mathcal{O}_{#1}^{'}}\xspace}                    

\newcommand{\nospaceunit}[1]{\ensuremath{\text{#1}}}       
\newcommand{\aunit}[1]{\ensuremath{\text{\,#1}}}       

\newcommand{\tev}{\aunit{Te\kern -0.1em V}\xspace}
\newcommand{\gev}{\aunit{Ge\kern -0.1em V}\xspace}
\newcommand{\mev}{\aunit{Me\kern -0.1em V}\xspace}
\newcommand{\kev}{\aunit{ke\kern -0.1em V}\xspace}
\newcommand{\ev}{\aunit{e\kern -0.1em V}\xspace}
 
\newcommand{\mevc}{\ensuremath{\aunit{Me\kern -0.1em V\!/}c}\xspace}
\newcommand{\gevc}{\ensuremath{\aunit{Ge\kern -0.1em V\!/}c}\xspace}
\newcommand{\mevcc}{\ensuremath{\aunit{Me\kern -0.1em V\!/}c^2}\xspace}
\newcommand{\gevcc}{\ensuremath{\aunit{Ge\kern -0.1em V\!/}c^2}\xspace}

\def\mum  {\ensuremath{\,\upmu\nospaceunit{m}}\xspace}

\def\fb   {\ensuremath{\aunit{fb}}\xspace}
\def\invfb   {\ensuremath{\fb^{-1}}\xspace}

\def\ps   {\ensuremath{\aunit{ps}}\xspace}

\def\gsim{{~\raise.15em\hbox{$>$}\kern-.85em
          \lower.35em\hbox{$\sim$}~}\xspace}
\def\lsim{{~\raise.15em\hbox{$<$}\kern-.85em
          \lower.35em\hbox{$\sim$}~}\xspace}

\def\sPlot{\mbox{\em sPlot}\xspace}

\def\pt         {\ensuremath{p_{\mathrm{T}}}\xspace}

\def\ptot       {\ensuremath{p}\xspace}

\def\evtgen     {\mbox{\textsc{EvtGen}}\xspace}

\def\geant      {\mbox{\textsc{Geant4}}\xspace}

\def\photos     {\mbox{\textsc{Photos}}\xspace}

\def\pythia     {\mbox{\textsc{Pythia}}\xspace}

\def\tell1  {TELL1\xspace}
\def\ukl1   {UKL1\xspace}

\newcommand{\lhcborcid}[1]{\href{https://orcid.org/#1}{\hspace*{0.1em}\raisebox{-0.45ex}{\includegraphics[width=1em]{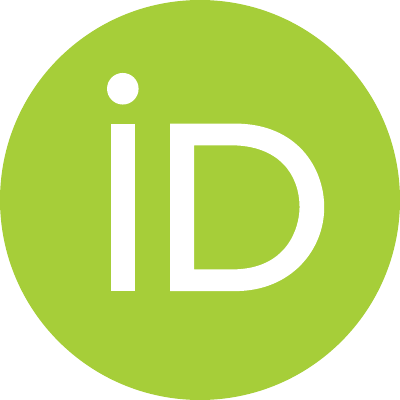}}}}

\usepackage{cite} 
\usepackage{mciteplus}

\usepackage{lineno}  
\usepackage{xspace} 
\usepackage{graphicx}  
\usepackage{color}
\usepackage{colortbl}
\graphicspath{{./figs/}} 

\usepackage{amsmath} 

\usepackage{amssymb}
\usepackage{amsfonts}
\usepackage{upgreek} 

\usepackage{enumerate}

\def\RResultDDSim {{\ensuremath{0.181 \pm 0.025}\xspace}}
\def\RResultLLSim {{\ensuremath{0.172 \pm 0.047}\xspace}}

\def\RResultTotSim {{\ensuremath{0.179 \pm 0.022 \pm 0.014}}}
\def\AbsBrResultTotSim {{\ensuremath{(1.24 \pm 0.15\pm 0.10 \pm 0.28 \pm 0.11)\times 10^{-4}}\xspace}}

\def\LbtoDDL {{\ensuremath{\Lb\to\Dp\Dm\Lz}\xspace}} 
\def\BtoDDK {{\ensuremath{\Bz\to\Dp\Dm\KS}\xspace}} 
\def\KsDecay {{\ensuremath{\KS\to\pip\pim}\xspace}} 
\def\LzDecay {{\ensuremath{\Lz\to\proton\pim}\xspace}} 
\def\DpDecay {{\ensuremath{\Dp\to\Km\pip\pip}\xspace}} 

\def\catLL {{\emph{LL}\xspace}} 
\def\catDD {{\emph{DD}\xspace}}

\def\calR {{\ensuremath{{\mathcal{R}}}\xspace}} 

\begin{document}

\renewcommand{\thefootnote}{\fnsymbol{footnote}}
\setcounter{footnote}{1}

\begin{titlepage}
\pagenumbering{roman}

\vspace*{-1.5cm}
\centerline{\large EUROPEAN ORGANIZATION FOR NUCLEAR RESEARCH (CERN)}
\vspace*{1.5cm}
\noindent
\begin{tabular*}{\linewidth}{lc@{\extracolsep{\fill}}r@{\extracolsep{0pt}}}
\ifthenelse{\boolean{pdflatex}}
{\vspace*{-1.5cm}\mbox{\!\!\!\includegraphics[width=.14\textwidth]{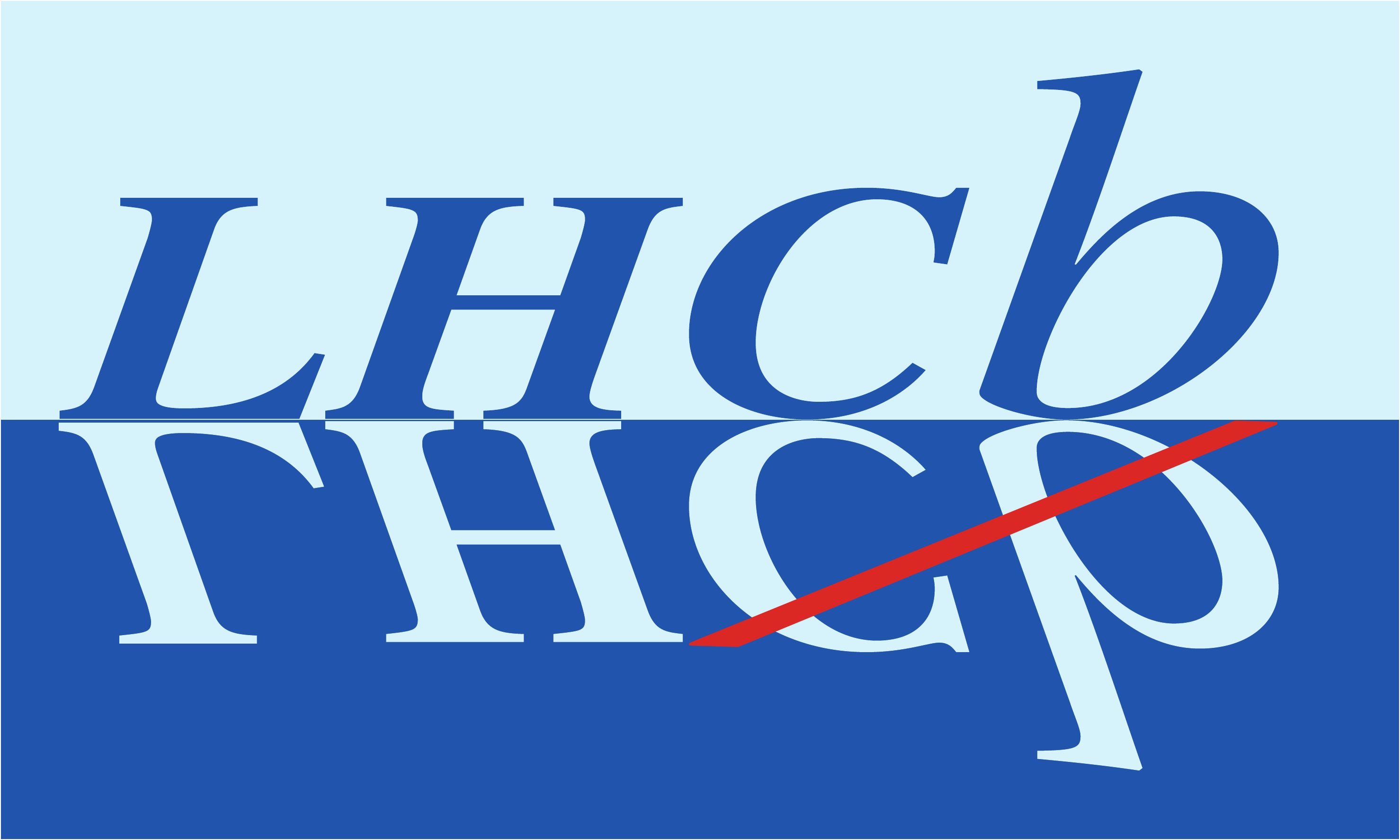}} & &}%
{\vspace*{-1.2cm}\mbox{\!\!\!\includegraphics[width=.12\textwidth]{figs/lhcb-logo.eps}} & &}%
\\
 & & CERN-EP-2024-041 \\  
 & & LHCb-PAPER-2023-042 \\  
 & & July 21, 2024 \\ 
 & & \\
\end{tabular*}

\vspace*{3.5cm}

{\normalfont\bfseries\boldmath\huge
\begin{center}
  \papertitle 
\end{center}
}

\vspace*{1.0cm}

\begin{center}
\paperauthors\footnote{Authors are listed at the end of this paper.}
\end{center}

\vspace{\fill}

\begin{abstract}
  \noindent
  The $\LbtoDDL$ decay is observed for the first time using proton-proton collision data collected by the \lhcb experiment at a center-of-mass energy of $13\tev$, corresponding to an integrated luminosity of $5.3\invfb$. Using the $\BtoDDK$ decay as a reference channel, the product of the relative production cross-section and decay branching fractions is measured to be
  \[
  \calR=\frac{\sigma_{\Lb}}{\sigma_{\Bz}} \times \frac{\BR(\LbtoDDL)}{\BR(\BtoDDK)}=\RResultTotSim,
  \]
  where the first uncertainty is statistical and the second is systematic.
  The known branching fraction of the reference channel, $\BR(\BtoDDK)$, and the cross-section ratio, $\sigma_{\Lb}/\sigma_{\Bz}$, previously measured by LHCb are used to derive the branching fraction of the $\LbtoDDL$ decay
  \[
  \BF(\LbtoDDL)=\AbsBrResultTotSim,
  \]
  where the third and fourth contributions are due to uncertainties of $\BR(\BtoDDK)$ and $\sigma_{\Lb}/\sigma_{\Bz}$, respectively.
  Inspection of the $\Dp\Lz$ and $\Dp\Dm$ invariant-mass distributions suggests a rich presence of intermediate resonances in the decay.
  The $\Lb\to\Dstarp\Dm\Lz$ decay is also observed for the first time as a partially reconstructed component in the $\Dp\Dm\Lz$ invariant mass spectrum.
\end{abstract}

\vspace*{0.1cm}

\begin{center}
  Published in JHEP 07 (2024) 140
\end{center}

\vspace{\fill}

{\footnotesize 
\centerline{\copyright~\papercopyright. \href{\paperlicenceurl}{\paperlicence}.}}
\vspace*{2mm}

\end{titlepage}


\newpage
\setcounter{page}{2}
\mbox{~}


\renewcommand{\thefootnote}{\arabic{footnote}}
\setcounter{footnote}{0}

\cleardoublepage


\pagestyle{plain} 
\setcounter{page}{1}
\pagenumbering{arabic}


\section{Introduction}
\label{sec:Introduction} 
Beauty baryons containing at least one $b$ quark have rich phenomenology, but most of them have not been fully explored yet.
The large sample of beauty baryons produced at the Large Hadron Collider (LHC) provides an unprecedented opportunity to enrich our knowledge of beauty baryons and better understand strong and weak interactions.
The $b\to c$ transition underlies the main decay modes of beauty baryons. 
From the late 1990s to the early 2010s, beauty baryon decays involving charmed baryons in the final state were almost exclusively known and long considered dominant~\cite{ALEPH:1996qft,CDF:2006vnk,LHCb-PAPER-2011-016}. 
Since 2014, however, decays with a $c$ quark hadronizing into a meson instead of a baryon have been commonly discovered by the LHCb experiment, with sizable branching fractions as well~\cite{LHCb-PAPER-2013-056,LHCb-PAPER-2014-020,LHCb-PAPER-2021-040}.
Among these, $b\to\ccbar s$ transitions provided the place to observe the first decays into systems made of a baryon and a charmonium resonance~\cite{D0:2007gjs,CDF:2009sbo,LHCb-PAPER-2015-032,LHCb-PAPER-2017-041}. 
For such final states, a significant contribution due to intermediate pentaquark states has been established~\cite{LHCb-PAPER-2022-031,LHCb-PAPER-2019-014,LHCb-PAPER-2017-043,LHCb-PAPER-2015-029}. 
Another class of decays that originate from $b\to\ccbar s$ transitions is the decay of a $\Lb$ baryon into a charmed baryon and an anti-charmed meson~\cite{LHCb-PAPER-2014-002}, which allows a precise measurement of the $\Lb$ baryon mass to be performed due to the limited phase space of the reaction.

\begin{figure}[!htbp]
\begin{center}
   \includegraphics[width=0.405\textwidth]{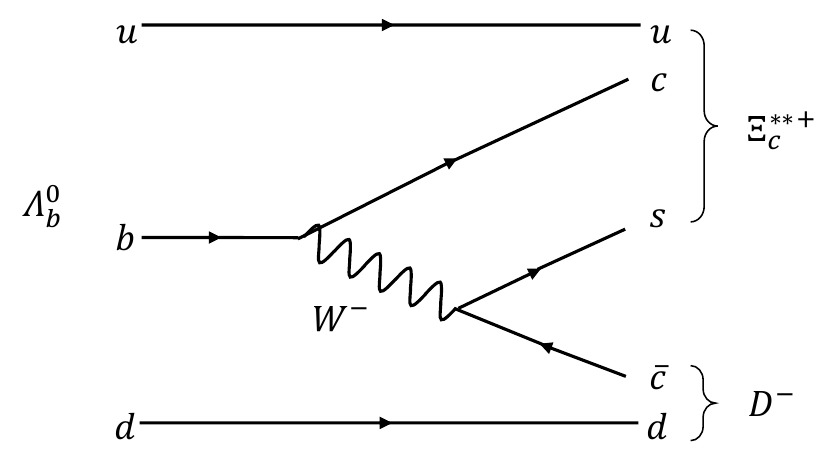}\\
   \includegraphics[width=0.495\textwidth]{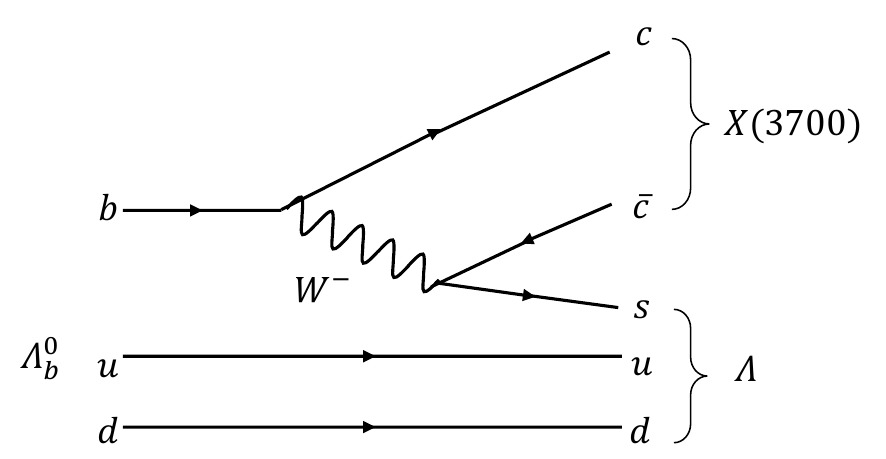}
   \includegraphics[width=0.495\textwidth]{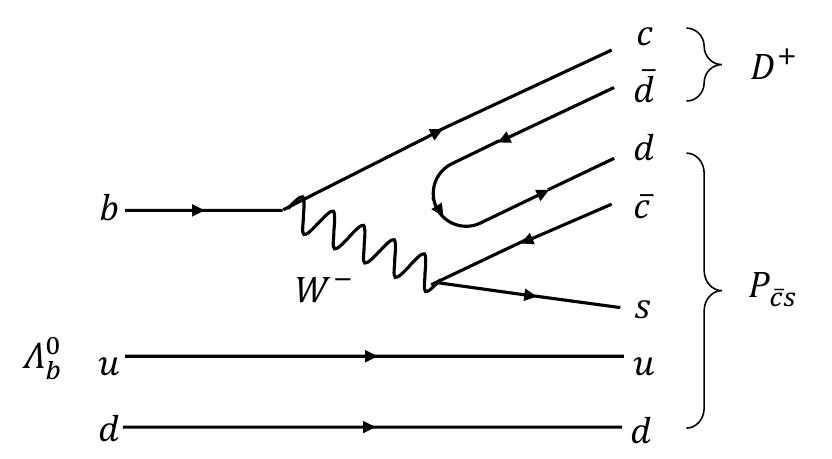}
\end{center}
\caption{Example diagrams resulting in intermediate states that might contribute to the $\LbtoDDL$ decay.}
\label{fig:feynman_diagram}
\end{figure}

This paper presents the first observation of the $\LbtoDDL$ decay,\footnote{The inclusion of charge-conjugate modes is always implied.
} referred to hereafter as the signal channel. 
This decay is of interest as it can proceed via two types of two-body intermediate states: one involving a charmed baryon and an anti-charm meson, and the other through charmonium(-like) states.
As discussed in Ref.~\cite{Wei:2021usz}, a $\Dp\Dm$ bound state can be produced near the $\Dp\Dm$ mass threshold, and open-charmed pentaquark states with quark content $\cquarkbar\squark\uquark\dquark\dquark$ might be present in the $\Dm\Lz$ final state.
Figure \ref{fig:feynman_diagram} shows the example diagrams resulting in these intermediate states that might contribute to the $\LbtoDDL$ decay, which originates from a $b\to\ccbar s$ transition. These include the $\Xic^{**+}$ baryon decaying to $\Dp\Lz$, the $X(3700)$ state decaying to $\Dp\Dm$ and pentaquark states $P_{\cquarkbar s}$ decaying to $\Dm\Lz$.

There is no previous experimental information about the signal channel $\Lambda_b^0 \to D^+ D^- \Lambda$. In the context of three-body B-meson decays, the doubly Cabibbo-suppressed decay $B^0 \to D^0 K^+ \pi^-$, for example, has a branching fraction on the order of $10^{-6}$~\cite{PDG2022}, the color-suppressed decay $B \to D^+ D^- K$ is around $10^{-4}$~\cite{BaBar:2010tqo}, and the Cabibbo-allowed decay $B^0 \to D^{*}(2010)^- \pi^+ \pi^0$ is at the $10^{-2}$ level~\cite{ARGUS:1990jet}.
Therefore, it can be expected that the branching fraction of $\Lambda_b^0 \to D^+ D^- \Lambda$, also a color-suppressed decay, could be comparable to these orders of magnitude.

This analysis is performed using proton-proton ($pp$) collision data collected by the \lhcb experiment from 2016 to 2018 at a center-of-mass energy of $13\tev$, corresponding to an integrated luminosity of $5.3\invfb$.
The branching fraction \BF of the $\LbtoDDL$ decay is determined by taking the $\BtoDDK$ decay as a reference channel. The latter has a decay topology similar to that of the signal channel, allowing for the cancellation of some systematic uncertainties in the branching fraction measurement. 
The primary measured quantity is the product of the relative production cross-section between $\Lb$ and $\Bz$ hadrons and  their decay branching fraction ratio, defined as
\begin{equation}
    \mathcal{R} \equiv \frac{\sigma_{\Lb}}{\sigma_{\Bz}} \times \frac{\BR(\LbtoDDL)}{\BR(\BtoDDK)},
\end{equation}
where $\sigma_{\Lb}$ ($\sigma_{\Bz}$) is the production cross-section of the $\Lb$ ($\Bz$) hadron, and \BF denotes the branching fractions. 
The branching fraction of the $\LbtoDDL$ decay is then derived using the value of the cross-section ratio, $\sigma_{\Lb}/\sigma_{\Bz}$, previously measured at the LHCb experiment~\cite{LHCb-PAPER-2018-050} and the known branching fraction of the reference channel~\cite{PDG2022}.

\section{Detector and data samples}
\label{sec:Detector}

The \lhcb detector~\cite{LHCb-DP-2008-001,LHCb-DP-2014-002} is a single-arm forward
spectrometer covering the \mbox{pseudorapidity} range $2<\eta <5$,
designed for the study of particles containing \bquark or \cquark
quarks. The detector includes a high-precision tracking system
consisting of a silicon-strip vertex detector surrounding the $pp$
interaction region, a large-area silicon-strip detector located
upstream of a dipole magnet with a bending power of about
$4{\mathrm{\,T\,m}}$, and three stations of silicon-strip detectors and straw
drift tubes placed downstream of the magnet.
The tracking system provides a measurement of the momentum, \ptot, of charged particles with
a relative uncertainty that varies from 0.5\% at low momentum to 1.0\% at 200\gevc.
The minimum distance of a track to a primary $pp$ collision vertex (PV), the impact parameter, 
is measured with a resolution of $(15+29/\pt)\mum$,
where \pt is the component of the momentum transverse to the beam, in\,\gevc.
Different types of charged hadrons are distinguished using information
from two ring-imaging Cherenkov detectors. 
Photons, electrons and hadrons are identified by a calorimeter system consisting of
scintillating-pad and preshower detectors, an electromagnetic
and a hadronic calorimeter.

The online event selection is performed by a trigger, 
which consists of a hardware stage, based on information from the calorimeter and muon
systems, followed by a software stage, which applies a full event
reconstruction. At the hardware trigger stage, events are required to have a hadron
with a transverse energy above $3.5\gev$ in the calorimeters.
The software trigger requires a two-, three- or four-track
secondary vertex with a significant displacement from any PV. At least one charged particle from the secondary vertex
must have a $\pt > 1.6\gevc$ and be
inconsistent with originating from any PV.

Simulation samples are required to model the effects of the detector acceptance and the
imposed selection requirements, and to study the invariant mass distributions of reconstructed $\LbtoDDL$ and $\BtoDDK$ decays. 
The simulation samples are generated with $\Lb$ or $\Bz$ signal decays uniformly distributed in the $\Dp\Dm\Lz$ or $\Dp\Dm\KS$ phase space.
In the simulation, $pp$ collisions are generated using
\pythia~\cite{Sjostrand:2007gs,*Sjostrand:2006za} 
with a specific \lhcb configuration~\cite{LHCb-PROC-2010-056}.
Decays of unstable particles
are described by \evtgen~\cite{Lange:2001uf}, in which final-state
radiation is generated using \photos~\cite{davidson2015photos}.
The interaction of the generated particles with the detector, and its response,
are implemented using the \geant
toolkit~\cite{Allison:2006ve, *Agostinelli:2002hh} as described in
Ref.~\cite{LHCb-PROC-2011-006}.

\section{Selection of candidates}
\label{sec:Selections}

In the offline reconstruction,  charged tracks of kaons, pions and protons are combined to form $\DpDecay$, $\LzDecay$ and $\KsDecay$ decays, which are then used to build the $\LbtoDDL$ and $\BtoDDK$ candidates. It is worth noting that the $\Lz$ and $\KS$ candidates (collectively referred to as the $\Vzero$ particles) are reconstructed according to two different categories. In the $\catLL$ category, they decay closely enough to the PV such that the final-state particles are reconstructed using the full tracking system. In the $\catDD$ category, the $\Vzero$ particles decay downstream of the vertex detector.
Consequently, the $\Vzero$ candidates have better momentum, mass and vertex resolution in the $\catLL$ category than in the $\catDD$ one, while the $\Vzero$ candidates in the $\catDD$ category have higher efficiency. 

A series of selection criteria are applied to the formed beauty-hadron candidates to suppress the background. The selection starts with loose requirements on kinematics, particle identification (PID) and variables that exploit the relatively long lifetime of beauty hadrons. 
Firstly, all final-state particles are required to have good track-fit quality and be displaced from any PV. 
Furthermore, these particles must have $\pt > 0.1\gevc$ and $p>2\gevc$, and be identified with a high significance as a pion, a kaon or a proton, using information from the tracking system and PID detectors. 
The scalar sum of $\pt$ of the $\Dpm$ candidates' decay products must be larger than $1.8\gevc$, with at least one of these having $\pt>0.5\gevc$ and $p>5\gevc$. 
The $\Dp$ and $\Vzero$ candidates are required to have a good-quality decay vertex that is significantly displaced from any PV.
The invariant mass of $\Dp$ candidates should lie within $25 \mevcc$ of the known mass~\cite{PDG2022}.
Moreover, $\Vzero$ candidates should have $\pt>250\mevc$ and an invariant mass within $6.6\,(5.4) \mevcc$ of the known masses~\cite{PDG2022} for the $\catDD$ ($\catLL$) category. 
The beauty hadron candidate formed by combining the $\Dpm$ and $\Vzero$ hadrons must have a good decay vertex displaced from its associated PV, which is defined as the PV that is the most compatible with the flight direction of the beauty candidate.
Additionally, its decay time is required to be greater than $0.2\ps$, and its momentum must point back to the associated PV. The final-state tracks of the $\Lb$ baryon must have a scalar $\pt$ sum larger than $5\gevc$, and at least one of the final-state particles must have $\pt$ larger than $1.7\gevc$ and $p$ larger than $10\gevc$, and at least two of them must have $\pt$ larger than $0.5\gevc$ and $p$ larger than $5\gevc$. 
A kinematic fit to the whole decay chain is performed with the $\Dp$ and $\Vzero$ constrained to their known masses~\cite{PDG2022} and the beauty-candidate momentum constrained to point back to the associated PV~\cite{Hulsbergen:2005pu}, which helps to improve the beauty-candidate mass resolution.

To further suppress the combinatorial background, 
a Boosted Decision Tree (BDT)\mbox{~\cite{Breiman,AdaBoost}} classifier implemented in the TMVA toolkit~\cite{Hocker:2007ht,*TMVA4} is employed.
Given the similar topology of the $\LbtoDDL$ and $\BtoDDK$ decays, the BDT classifier is trained using $\LbtoDDL$ samples and is applied to both decay modes, helping to cancel certain systematic uncertainties associated with the efficiency determination in the branching fraction measurement.
The BDT classifier training uses candidate $\Lb$ decays in the high mass sideband ($5.80<m(\Dp\Dm\Lz)<6.15\gevcc$) in data, which serves as a proxy for the background, and simulated $\LbtoDDL$ signal decays.
The training is performed separately for the $\catLL$ and $\catDD$ categories.
The BDT classifier combines variables including the beauty-hadron decay topology, PID information of final-state particles and kinematic properties of the $\Lb$, $\Dpm$, $\Lz$ and final-state candidates, to discriminate between the signal and the background. 
The optimal requirement on the BDT response is determined by maximising the Punzi figure of merit $\epsilon/(\frac{5}{2}+\sqrt{B})$~\cite{Punzi:2003bu}, where $\epsilon$ is the signal efficiency and $B$ is the background yield in the data signal region.
The signal region is chosen as $\pm25\mevcc$ around the known $\Lb$ mass~\cite{PDG2022}.
The signal efficiency $\epsilon$ is estimated from simulated samples, and the background yield $B$ is estimated by a linear extrapolation of the yield in the high-mass data sideband to the signal region. 
The requirements on the BDT response are optimized separately for the signal and the reference channels but are found to be similar.
The optimal requirement on the BDT response rejects about 99\% of the combinatorial background while maintaining a signal efficiency of about 80\%.

The background from the $\KS\to\pip\pim$ decay, which may be reconstructed as the $\Lz\to\proton\pim$ decay due to misidentification of a pion as a proton, is investigated. 
This cross-feed background is suppressed by imposing a tight PID requirement on the proton track if the $p\pim$ invariant mass under the $\pip\pim$ hypothesis is consistent with the $\KS$ meson mass. 
The remaining cross-feed background in the $\BtoDDK$ decay is found to be negligible.
The background from \mbox{$\Lb\to\Dp\Kp\pim\pim\Lz$,} $\Lb\to\Km\pip\pip\Dm\Lz$ and $\Lb\to\Km\pip\pip\Kp\pim\pim\Lz$ decays is referred to hereafter as the non-double-charm (NDC) background. This background involves the same set of final-state particles as the signal decays but there is only one or no intermediate $\Dpm$ meson.
The NDC background is largely suppressed by the imposed requirement on the $\Dpm$ invariant mass. 
However, any residual NDC background can produce a peak in the $\Dp\Dm\Lz$ invariant mass distribution, and its yield is estimated based on the number of candidates in the $\Dpm$ mass sidebands, positioned within a range of $35$ to $90$ $\mevcc$ from the known $\Dpm$ mass.
The fraction of the NDC background in the $\Lb$ sample is estimated from a fit to the $\Dp\Dm\Lz$ invariant mass distribution in the $\Dpm$ mass sidebands and extrapolating to the signal region. It is determined to be $(6\pm7)\%$ and $(6\pm3)\%$ for the $\catLL$ and $\catDD$ categories, respectively. Similar results are measured for the $\BtoDDK$ decay channel. 
This method aligns with that used in Refs.~\cite{LHCb-PAPER-2020-025,LHCb-PAPER-2022-019}.
The obtained NDC yield is then subtracted from the total beauty-hadron yield.

\section{Signal yield determination}
\label{sec:Fit}
\begin{figure}[t]
\begin{center}
   \includegraphics[width=0.496\textwidth]{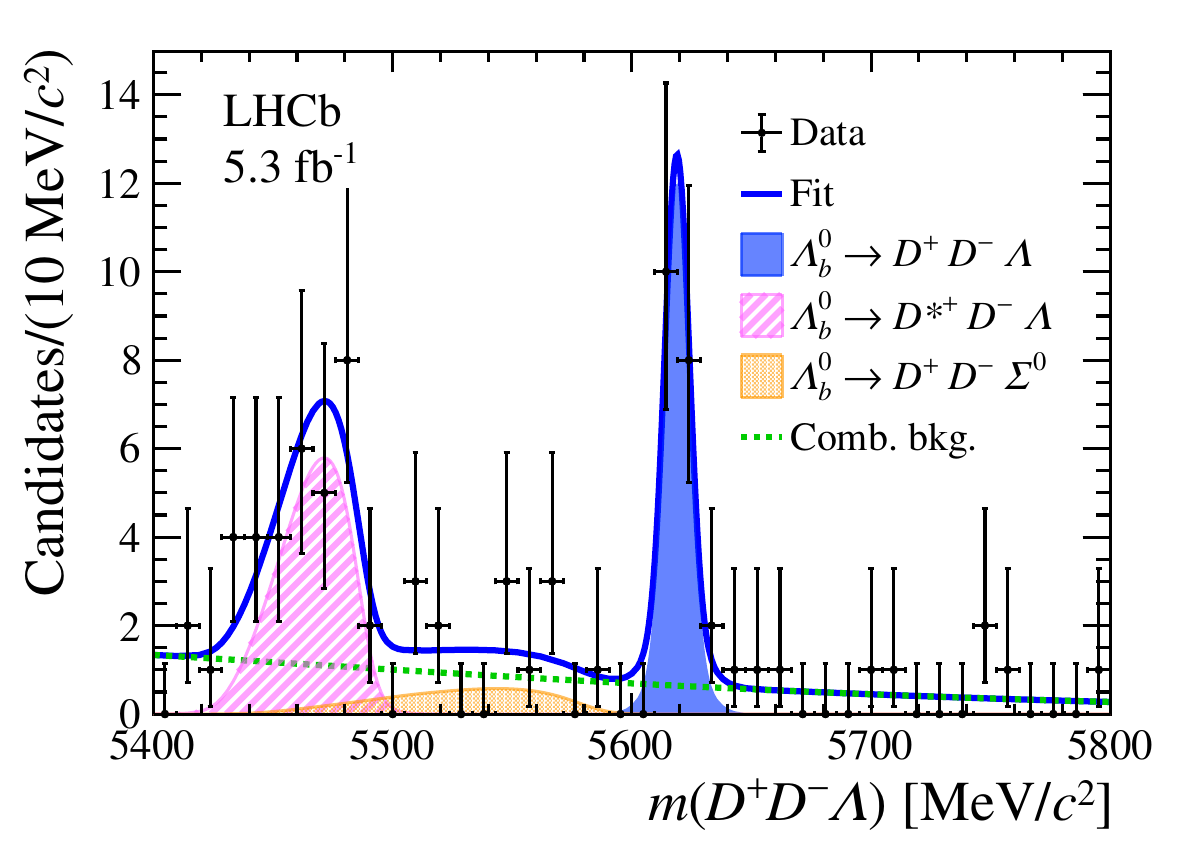}
   \includegraphics[width=0.496\textwidth]{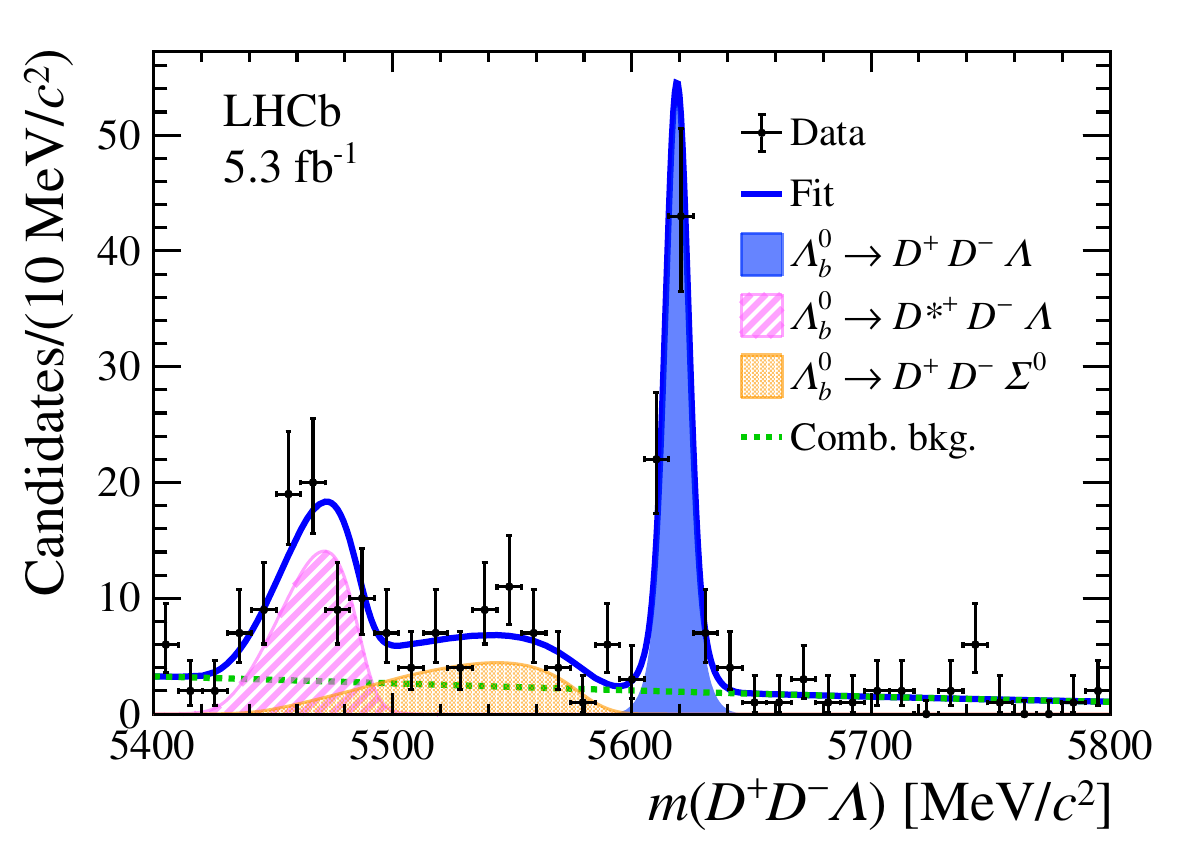}
\end{center}
\caption{
    Invariant mass distribution of $\LbtoDDL$ candidates near the signal region with the fit results superimposed, for the (left) $\catLL$ and (right) $\catDD$ categories, respectively. 
}
\label{fig:fitData_Lb}
\end{figure}

\begin{figure}[t]
\begin{center}
   \includegraphics[width=0.496\textwidth]{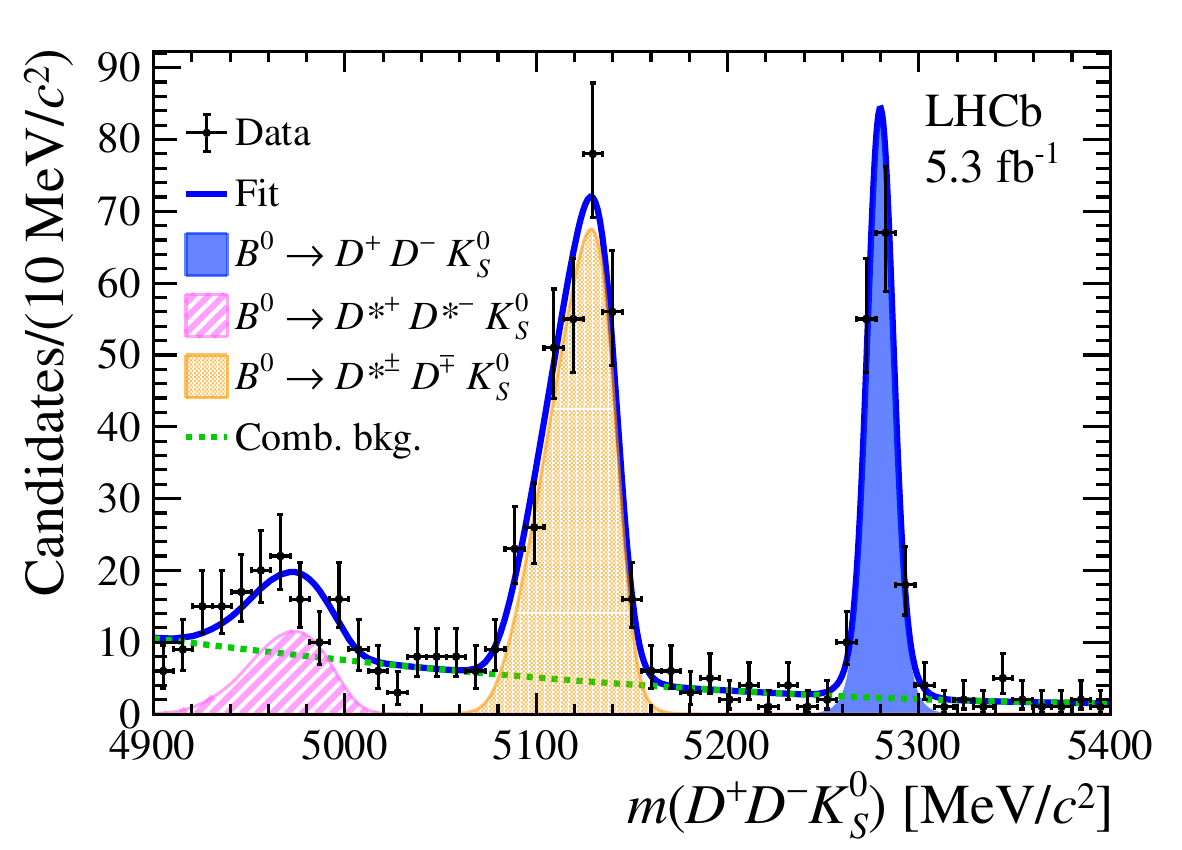}
    \includegraphics[width=0.496\textwidth]{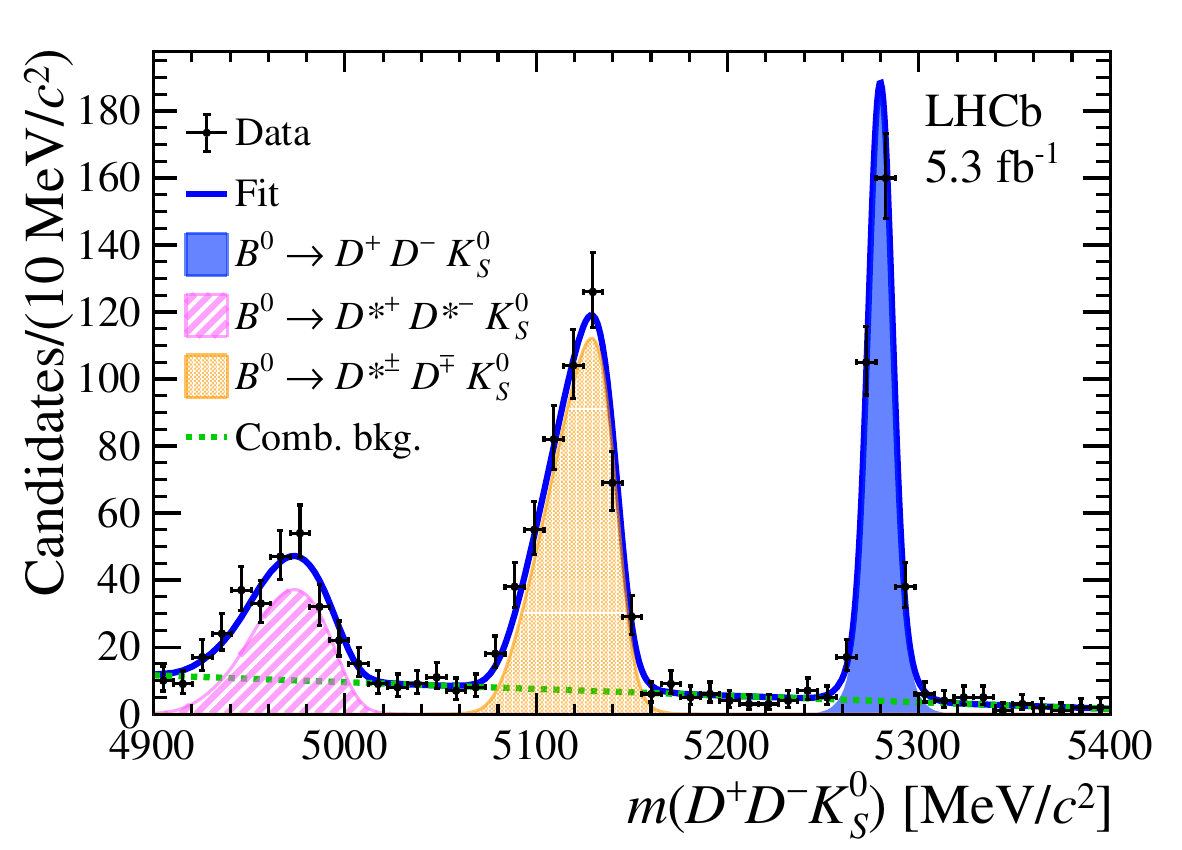}
\end{center}
\caption{
    Invariant mass distribution of $\BtoDDK$ candidates with the fit results superimposed, for the (left) $\catLL$ and (right) $\catDD$ categories, respectively.
}
\label{fig:fitData_B0}
\end{figure}

The invariant mass distributions of the $\Lb$ and $\Bz$ candidates after applying all the aforementioned selection requirements are shown in Figs.~\ref{fig:fitData_Lb} and \ref{fig:fitData_B0}, respectively. 
The distributions are characterized by three significant peaking structures, one for the signal decay and the other two for partially reconstructed beauty-hadron decays, and a wide smooth combinatorial background. 
These components are fit with an extended unbinned maximum-likelihood method to determine the $\Lb$ and $\Bz$ signal yields. 

The signal peak is described by the sum of two Gaussian functions sharing the same mean value. 
The common mean and the average resolution of the two Gaussian functions are free parameters in the fit, while the relative fractions of the two components and the ratio of their widths are fixed to the values obtained from the simulation. 
In Fig.~\ref{fig:fitData_Lb}, the peaking background closer to the signal peak originates from the \mbox{$\Lb\to\Dp\Dm\Sigmares^0(\to\Lz\gamma)$} decay where the $\gamma$ is not included. 
The left peaking background is consistent with $\Lb\to D^{*\pm}(\to\Dpm\piz)\Dmp\Lz$ decays with the $\piz$ not included.
In Fig.~\ref{fig:fitData_B0}, the partially reconstructed backgrounds in the $\BtoDDK$ mass spectrum are consistent with the $\Bz\to\Dstarp\Dstarm\KS$ and $\Bz\to\D^{*\pm}\Dmp\KS$ decays, where the $\piz$ in the $\D^{*\pm}\to\Dpm\piz$ decay is not included.
The shapes of the partially reconstructed backgrounds are obtained from a fast simulation~\cite{Cowan:2016tnm}, and convolved with the experimental resolution.
The combinatorial background is modelled empirically by slowly varying functions, namely the ARGUS function~\cite{ARGUS:1990hfq} for the $\LbtoDDL$ mode and a second-order polynomial for the $\BtoDDK$ mode, with all parameters free in the fit. 
The fit of $m(\Dp\Dm\Lz)$ distribution is performed over a larger range than what is shown in Fig.~\ref{fig:fitData_Lb} to better constrain the combinatorial background. 
The fit is performed simultaneously for the $\catLL$ and $\catDD$ samples, 
and the mean of the signal Gaussians is shared.

The fit results are shown in Fig.~\ref{fig:fitData_Lb} for $\LbtoDDL$ candidates and in Fig.~\ref{fig:fitData_B0} for the $\BtoDDK$ candidates.
After the subtraction of the aforementioned NDC background, the signal yields for the $\LbtoDDL$ and $\BtoDDK$ decays are determined to be $18\pm5$ and $128\pm12$ for the $\catLL$ category, $69\pm9$ and $296\pm18$ for the $\catDD$ category, respectively.
The statistical significance of the $\Lambda_b^0 \to D^+ D^- \Lambda$ decay evaluated using the likelihood-ratio test~\cite{Wilks:1938dza} exceeds 16 standard deviations ($\sigma$) after considering systematics.
Thus, this represents the first observation of the $\LbtoDDL$ decay mode. 
The left peaking structure in the $\Dp\Dm\Lz$ invariant mass distributions exhibits a statistical significance of $9\sigma$ using the same method as above. This structure is consistent with the $\Lb\to\Dstarp\Dm\Lz$ decay, being thus observed for the first time in the $\Dp\Dm\Lz$ invariant mass spectrum.

\section{Branching fraction measurement}
\label{sec:Br}
The branching fraction of the $\LbtoDDL$ decay is measured with respect to that of the $\BtoDDK$ decay as
\begin{equation}
    \begin{split}
        \mathcal{R} = \frac{N_\Lb^\mathrm{corr}}{N_\Bz^\mathrm{corr}} \times \frac{\BR(\KS \rightarrow \pip \pim)}{\BR(\Lz \rightarrow p \pim)},
    \label{equa:def_rela_br}
    \end{split}
\end{equation}
where $N^\mathrm{corr}_{X_b}$ is the efficiency corrected yield of the $X_b=\Lb\text{ or }\Bz$ decay.
The branching fractions 
$\BR(\Lz\to p\pim)$ and $\BR(\KS\to \pip\pim)$ are taken from Ref.~\cite{PDG2022}.
The efficiency corrected yield is determined as
\begin{equation}
    N^{\mathrm{corr}}_{X_b}=\sum_i \frac{w_i}{\epsilon_i},
    \label{equa:Ncorr_sig}
\end{equation}
where the sum runs over all the selected candidates in the signal or reference channel. 
The weight $w_{i}$ assigned to each candidate is calculated using the invariant mass fit result following the \sPlot technique~\cite{Pivk:2004ty}, which is used to statistically subtract the background contribution in the data sample.  
The $\epsilon_i$ denotes the total experimental efficiency, which is calculated as a function of the final-state phase-space coordinates, represented by the two-dimensional plot of the two-body invariant mass distributions $m(\Dp\Dm)$ and $m(\Dp\Vzero)$, and it is estimated for the $\catLL$ and $\catDD$ categories separately.
The efficiency function is determined using simulated samples, which have been calibrated such that the distributions of several key variables match the data, including the PID response, $\Lb$ kinematics and the total charged-track multiplicity. 
The relative variation of the total experimental efficiency across the phsp space is around 20\% for both the signal and reference channels in the $\catLL$ and $\catDD$ categories.

Thus, $\mathcal{R}$ is determined to be
$\RResultLLSim$ and $\RResultDDSim$ for the $\catLL$ and $\catDD$ categories, respectively, where the uncertainties are statistical.

\section{Systematic uncertainties}
\label{sec:Systematics}

The systematic uncertainties on the branching fraction measurement,
apart from external inputs, are generally related to either the signal yields or the efficiencies. 
Due to the similar topology of the signal and reference decays, many sources of systematic uncertainties are either cancelled or largely suppressed in the ratio of branching fractions. 
The remaining systematic uncertainties are outlined below and summarized in Table~\ref{table:uncertainty}.

The determination of the signal yields in the $\Lb$ and $\Bz$ decay channels is affected by the choice of the fit models for the various components, which in turn influences the branching fraction measurement. 
To test the signal model, alternative fits are performed by either fixing all the shape parameters to those obtained from the simulation or by using a Hypatia function~\cite{Santos:2013gra} with the parameters of mean and resolution floated in the fit. In order to study the systematic uncertainty associated with the combinatorial background modelling, the range of the beauty-hadron invariant mass used in the fit is modified. 
The relative deviation of the branching fraction obtained in these alternative fits from the baseline result is found to be 3.4\% at most, which is taken to be the systematic uncertainty due to the invariant mass modelling.

The distributions of the discriminating variables used in the BDT classifier are found to be consistent between the simulation and background-subtracted data of the reference channel.
Additionally, given the overlap of the distributions of discriminating variables between the signal and reference channels in simulation, no corresponding systematic uncertainties are quoted.
To validate the effect of the possible overtraining of the BDT classifier on the beauty-hadron invariant mass distribution, a fit is performed excluding the candidates used in the BDT training. The relative difference of the new branching fraction with respect to the baseline result is found to be 2.5\% and is taken as the corresponding uncertainty. 

The estimated NDC background fractions in the signal and reference channels suffer from the limited size of the data samples. These uncertainties are propagated to the branching fraction measurement causing a relative uncertainty of 4.5\%.

The efficiencies are estimated from simulated samples, so systematic contributions arise due to the limited size of the samples, and due to the imperfect simulation of the detector response and particle kinematics.
The relative uncertainty due to the limited size of the simulation samples is 3.9\%.
Corrections of the simulation samples for their different PID response and beauty-hadron kinematic distributions than in data are subject to uncertainties. 
Specifically, the systematic uncertainty associated with the corrections of the PID response is evaluated using alternative correction templates~\cite{LHCb-DP-2018-001} and measuring the relative change of efficiencies, which is found to be negligible. 
The uncertainties on the corrections of the \Lb distributions are propagated to the branching fraction measurement using pseudoexperiments. For every pseudoexperiment, the correction factor in each bin of the \Lb transverse momentum and rapidity is varied following a Gaussian distribution. The mean and width of the Gaussian function are taken as the baseline correction factor and its uncertainty.
Also, the branching fraction is calculated for each pseudoexperiment, and then a set of branching fractions is obtained, whose standard deviation is taken as the systematic uncertainty, calculated to be 1.5\%.

The hardware-trigger response is approximately modelled in the simulation, therefore, a data-driven calibration is performed to align the efficiency in the simulation with that in the data~\cite{MartinSanchez:1407893}. The uncertainties on the calibration factors are propagated to the branching fraction measurement, resulting in a relative uncertainty of 0.5\%.

The experimental efficiency is measured in bins of the final-state phase space. The choice of the binning scheme introduces another source of systematic uncertainty. The efficiency, recalculated using coarser bins, results in a relative change in the branching fraction of 2.6\%. This value is thus quoted as a systematic uncertainty.

\begin{table}[tb]
    \centering
    \caption{Summary of relative systematic uncertainties and uncertainties from external inputs on the branching fraction measurement. }
    \begin{tabular}{l c}
    \hline
    Source  & Relative uncertainty (\%) \\
    \hline
    Modelling of the invariant mass distribution & \phantom{0}3.4\\
    BDT classifier training & \phantom{0}2.5\\
    Subtraction of NDC background & \phantom{0}4.5\\
    Size of simulation samples & \phantom{0}3.9\\
    Corrections to simulation samples & \phantom{0}1.5\\
    Trigger efficiency & \phantom{0}0.5\\
    Binning scheme of two-body invariant mass  &  \phantom{0}2.6\\
    Branching fraction ratio of $\Lz\to p\pim$ and $\KS\to\pip\pim$ & \phantom{0}0.9\\
    Branching fraction of $\Bz\to\Dp\Dm\Kz$ & 23.1\\
    Beauty-hadron production cross-section ratio & \phantom{0}9.1\\
    \hline
    Total systematic uncertainty & \phantom{0}7.8\\
    Total uncertainty due to external inputs & 24.7\\
    \hline
    \end{tabular}
    \label{table:uncertainty}
\end{table}

Besides the contributions outlined above, there are also further uncertainties due to the external inputs. They come from the limited knowledge of the branching fractions of the intermediate decays $\Lz\to p\pim$, $\KS\to\pip\pim$ and that of the reference channel $\Bz\to\Dp\Dm\Kz$~\cite{BaBar:2010tqo}.
Benefiting from the precise measurement of branching fractions of $\Lz\to p\pim$ and $\KS\to\pip\pim$~\cite{PDG2022}, the uncertainties associated with these two intermediate decays are small, at 0.9\% relatively. 
However, experimental knowledge about the branching fraction of the reference channel $\Bz\to\Dp\Dm\Kz$ is quite limited, resulting in a large uncertainty of 23.1\%, which dominates the overall uncertainties. The precision of the branching fraction measurement of $\LbtoDDL$ can be greatly enhanced once the measurement of $\Bz\to\Dp\Dm\Kz$ is significantly improved.
Finally, another source is the uncertainty from the measurement of the beauty-hadron production cross-section ratio performed by \lhcb, which is $9.1\%$~\cite{LHCb-PAPER-2018-050}.
The total uncertainty due to the external inputs is 24.7\%, which is the dominant uncertainty compared with the statistical and systematic ones.

\section{Results and summary}
\label{sec:Results}
The product of the ratio of branching fractions and the corresponding beauty-hadron production cross-section is measured to be
\begin{equation*}
    \calR=\frac{\sigma_{\Lb}}{\sigma_{\Bz}} \times \frac{\BR(\LbtoDDL)}{\BR(\BtoDDK)}=\RResultTotSim,
\end{equation*}
where the first uncertainty is statistical and the second is systematic. 
This result represents a weighted average of the $\catLL$ and $\catDD$ samples, with weights calculated using the statistical uncertainties.

The relative production cross-section $\sigma_\Lb/\sigma_\Bz$ is equal to the ratio of hadronization fractions, $f_{\Lb}/f_{\Bz}$, which has been measured by \lhcb as a function of the \pt of the $\Lb$ baryon~\cite{LHCb-PAPER-2018-050}. Averaging over the \pt distribution of the \Lb baryon in data gives $\sigma_\Lb/\sigma_\Bz=0.541 \pm 0.048$.
Using the known branching fraction of $\BF(\Bz\to\Dp\Dm\Kz)$~\cite{BaBar:2010tqo} and assuming equal probabilities for observing a $\Kz$ meson as a $\KS$ or $\KL$ mass eigenstate,
the $\LbtoDDL$ absolute branching fraction is determined to be
\begin{equation*}
    \BR(\LbtoDDL)=\AbsBrResultTotSim,
\end{equation*}
where the first uncertainty is statistical, the second is systematic, and the third and fourth come from the uncertainties on the $\BtoDDK$ branching fraction and the beauty-hadron production cross-section ratio~\cite{LHCb-PAPER-2018-050}, respectively. 

\begin{figure}[!htbp]
\begin{center}
   \includegraphics[width=0.496\textwidth]{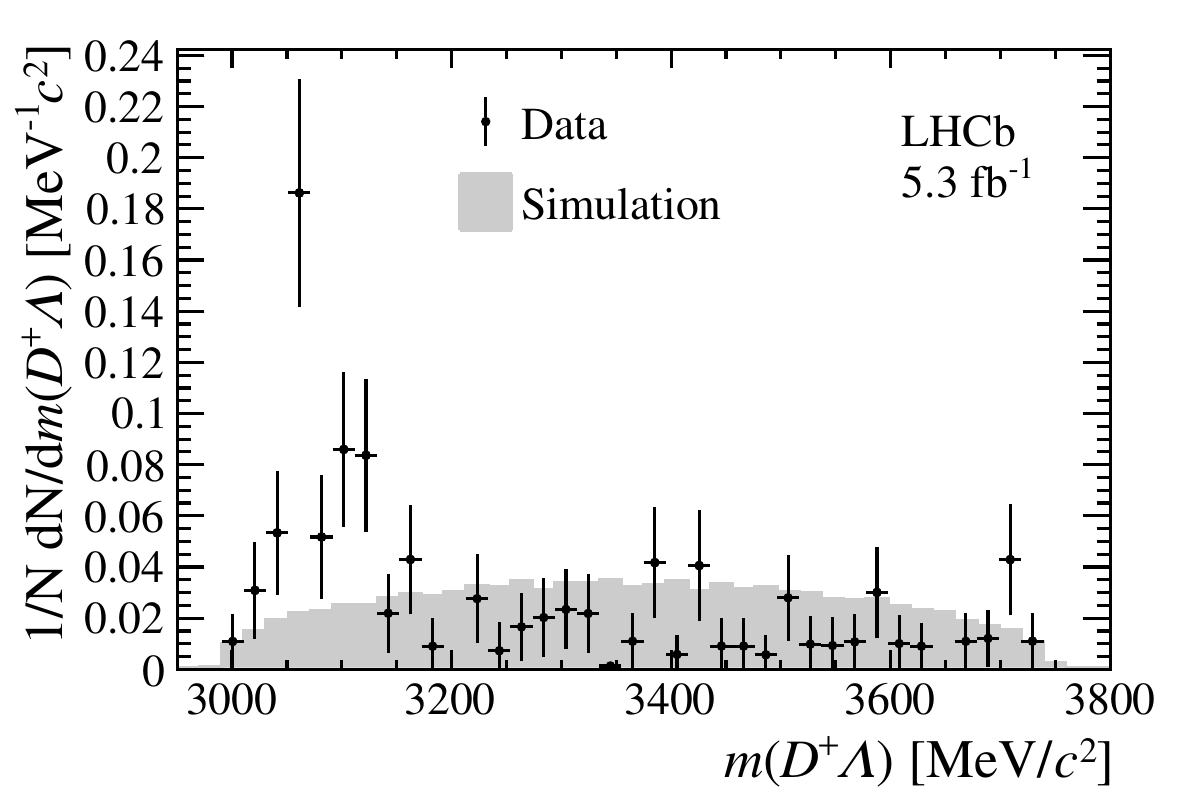}
   \includegraphics[width=0.496\textwidth]{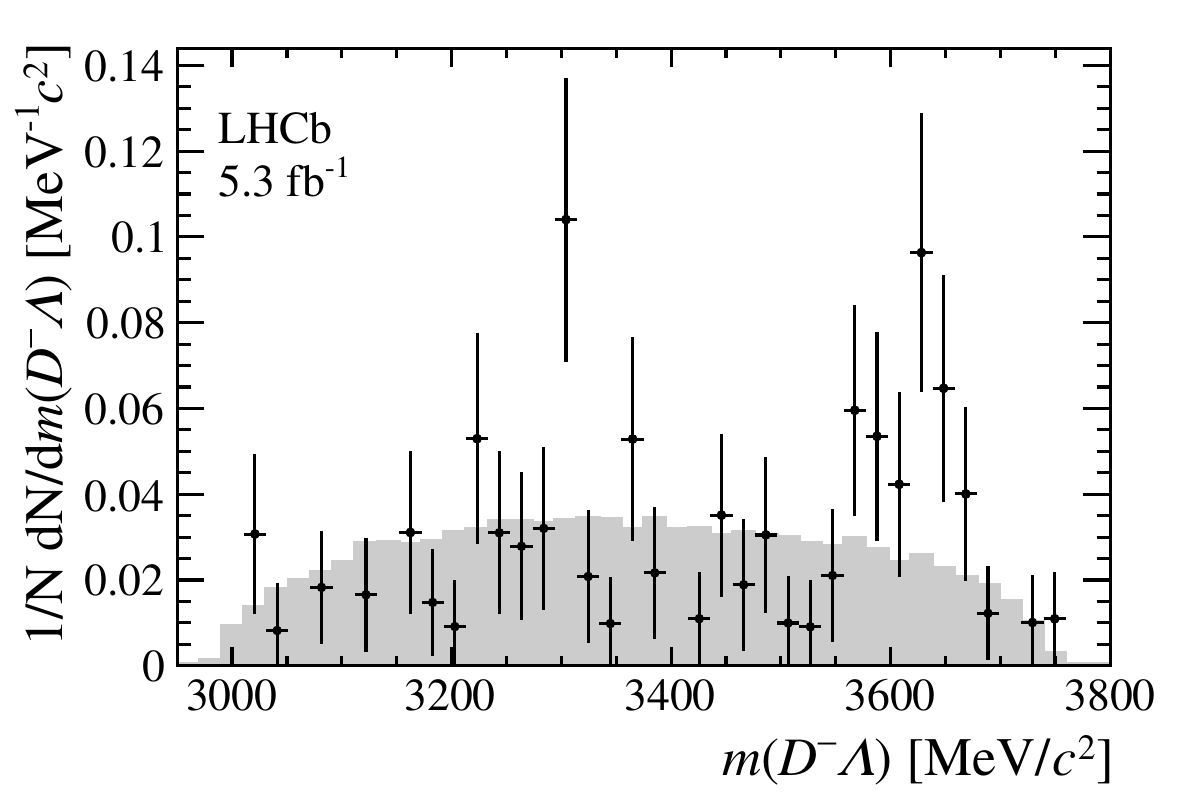}
   \includegraphics[width=0.496\textwidth]{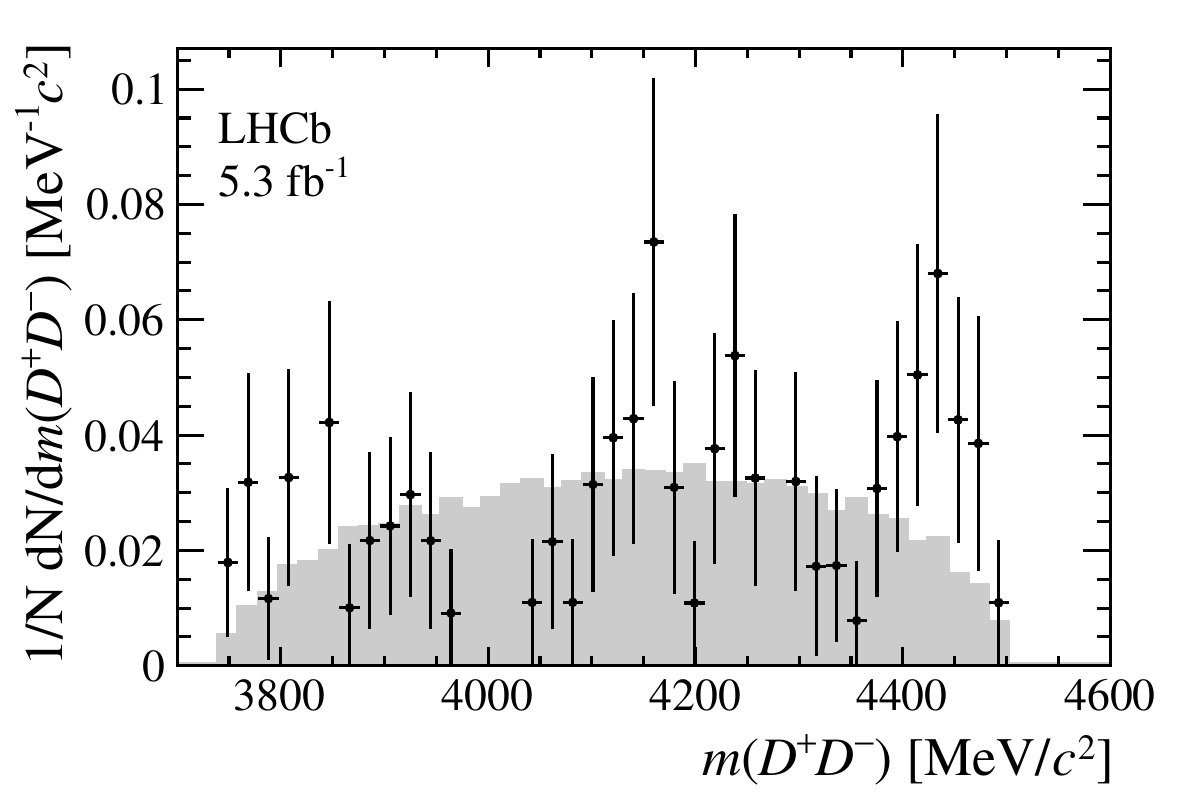}
\end{center}
\caption{
    Normalized and background-subtracted invariant mass distributions of the (left) $\Dp\Lz$, (right) $\Dm\Lz$ and (bottom) $\Dp\Dm$ systems in the $\LbtoDDL$ decay, compared with the distributions in simulation samples.
}
\label{fig:structure_sig}
\end{figure}

\begin{figure}[!htbp]
\begin{center}
   \includegraphics[width=0.496\textwidth]{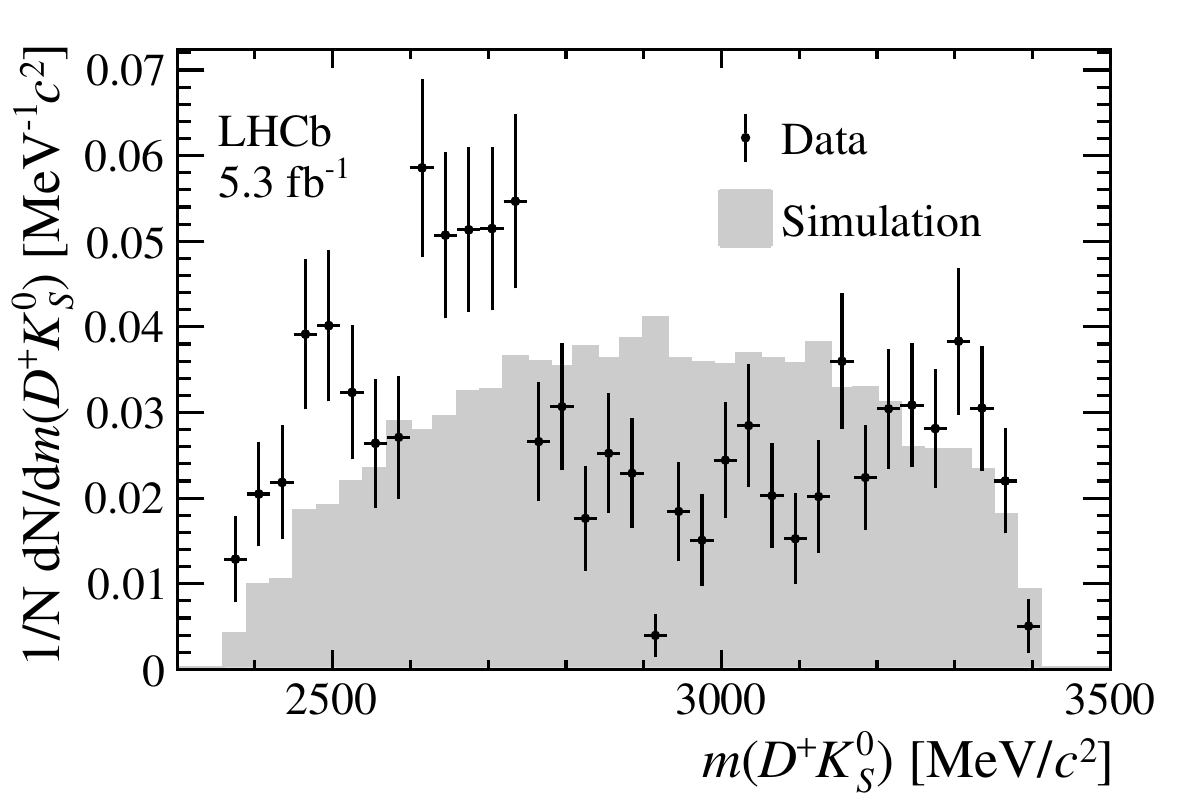}
   \includegraphics[width=0.496\textwidth]{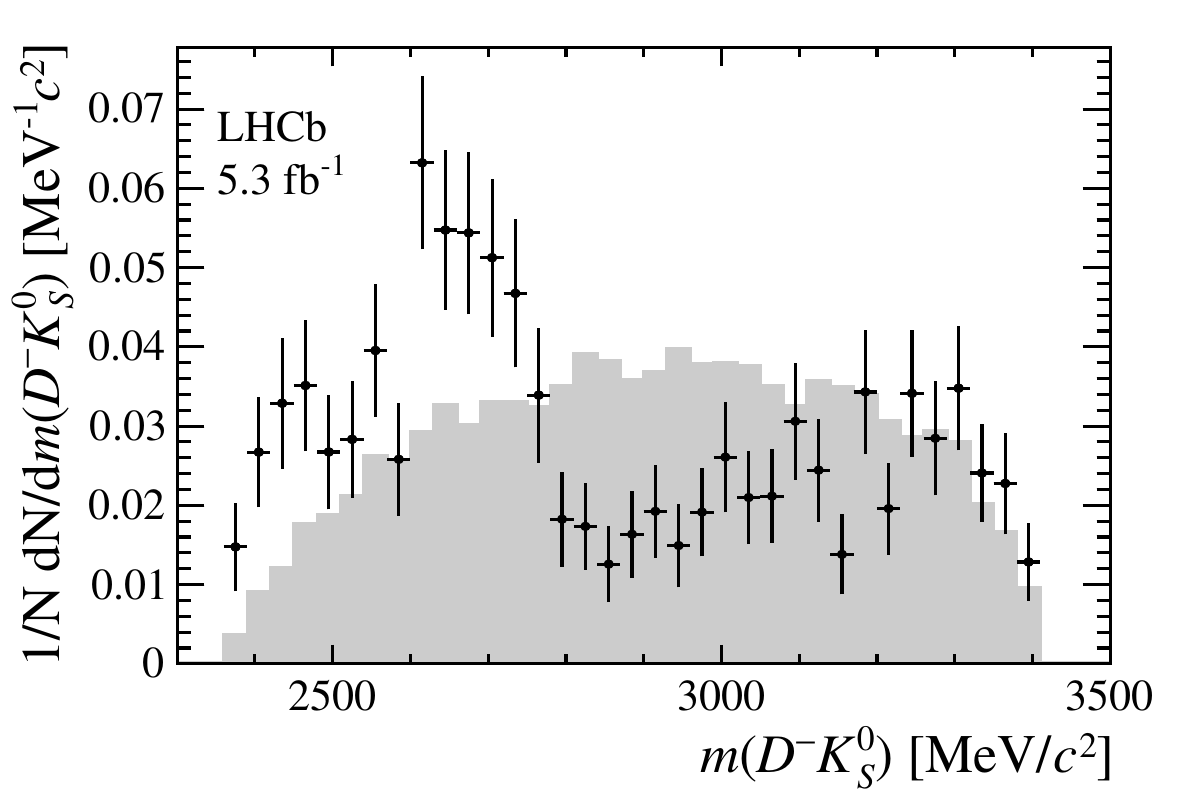}
   \includegraphics[width=0.496\textwidth]{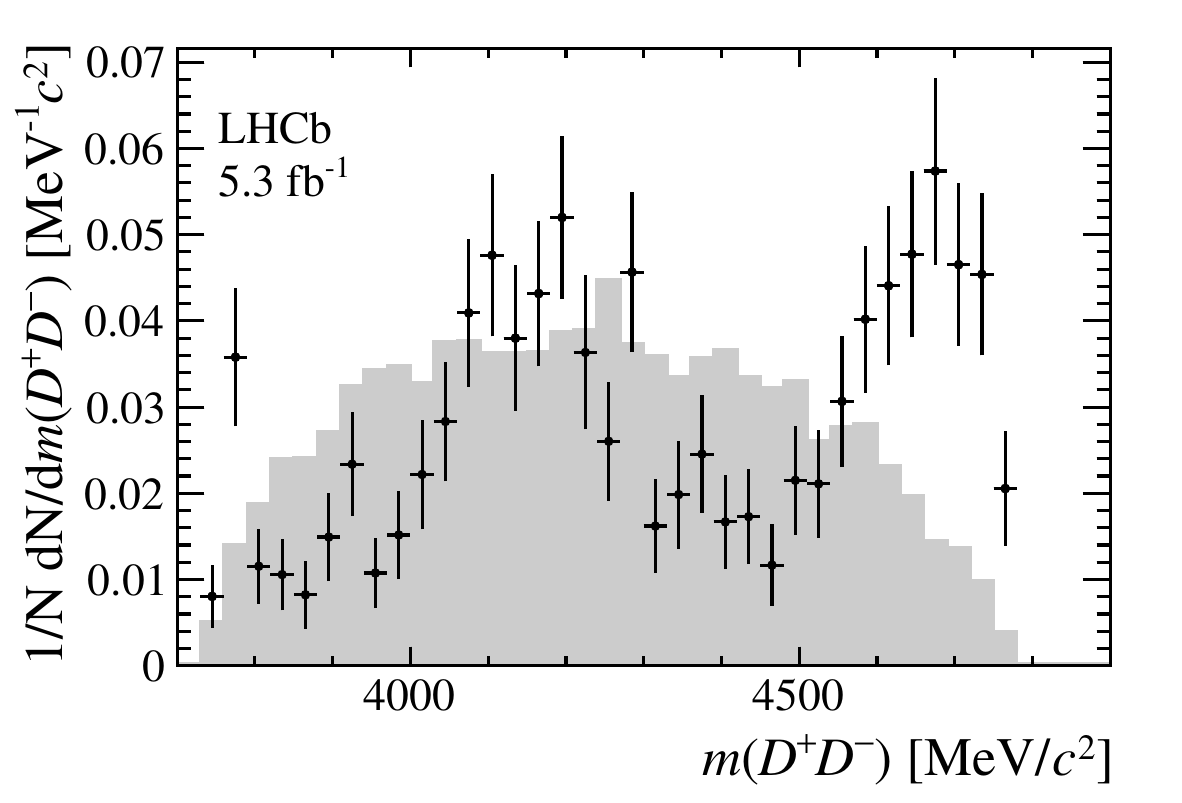}
\end{center}
\caption{
    Normalized and background-subtracted invariant mass distributions of the (left) $\Dp\KS$, (right) $\Dm\KS$ and (bottom) $\Dp\Dm$ systems in the $\BtoDDK$ decay, compared with the distributions in the simulation sample.
}
\label{fig:structure_con}
\end{figure}

As emphasized in Figs. \ref{fig:structure_sig} and \ref{fig:structure_con}, where the invariant mass distributions $m(\Dp\Dm)$, $m(\Dp\Vzero)$ and $m(\Dm\Vzero)$ are shown, these decays are particularly noteworthy for their potential in probing possible resonant structures. 
These invariant mass distributions, in which the combinatorial background is subtracted using the \sPlot technique, strongly depart from the phase space distributions obtained in simulation. 
Such deviations suggest the presence of a variety of resonant structures. For example, excited $\Xires_c^+$ states around $3055\mevcc$ are very likely to be included in the $m(\Dp\Lz)$ distribution, as shown in the top left plot of Fig.\ref{fig:structure_sig}. Additionally, the $\D\Dbar$ bound state $X(3700)$, predicted by the theory in Ref.~\cite{Wei:2021usz}, may exist in the $m(\Dp\Dm)$ distribution. However, as shown in the bottom plot of Fig.~\ref{fig:structure_sig}, the current statistics do not allow for a definitive conclusion. Furthermore, pentaquark states could potentially contribute to the $m(\Dm\Lz)$ distribution.
These open the door to future discoveries in upcoming studies with larger datasets.

In summary, the $\LbtoDDL$ decay is observed for the first time using proton-proton collision data collected by the \lhcb experiment at a center-of-mass energy of $13\tev$, corresponding to an integrated luminosity of $5.3\invfb$. 
The branching fraction of the $\LbtoDDL$ decay is determined, which is a key step towards probing potential intermediate resonant states.
The peaking structure in the $\Dp\Dm\Lz$ invariant mass distribution, which is consistent with the $\Lb\to\Dstarp\Dm\Lz$ decay, is also reported.
Future studies will be needed to measure the branching fraction of the $\Lb\to\Dstarp\Dm\Lz$ decay.
The indications of various resonant structures in the $\LbtoDDL$ decay also call for future exploration in similar decay channels, like the $\Lb\to\Dz\Dzb\Lz$ decay.

\clearpage
\section*{Acknowledgements}
\noindent We express our gratitude to our colleagues in the CERN
accelerator departments for the excellent performance of the LHC. We
thank the technical and administrative staff at the LHCb
institutes.
We acknowledge support from CERN and from the national agencies:
CAPES, CNPq, FAPERJ and FINEP (Brazil); 
MOST and NSFC (China); 
CNRS/IN2P3 (France); 
BMBF, DFG and MPG (Germany); 
INFN (Italy); 
NWO (Netherlands); 
MNiSW and NCN (Poland); 
MCID/IFA (Romania); 
MICINN (Spain); 
SNSF and SER (Switzerland); 
NASU (Ukraine); 
STFC (United Kingdom); 
DOE NP and NSF (USA).
We acknowledge the computing resources that are provided by CERN, IN2P3
(France), KIT and DESY (Germany), INFN (Italy), SURF (Netherlands),
PIC (Spain), GridPP (United Kingdom), 
CSCS (Switzerland), IFIN-HH (Romania), CBPF (Brazil),
and Polish WLCG (Poland).
We are indebted to the communities behind the multiple open-source
software packages on which we depend.
Individual groups or members have received support from
ARC and ARDC (Australia);
Key Research Program of Frontier Sciences of CAS, CAS PIFI, CAS CCEPP, 
Fundamental Research Funds for the Central Universities, 
and Sci. \& Tech. Program of Guangzhou (China);
Minciencias (Colombia);
EPLANET, Marie Sk\l{}odowska-Curie Actions, ERC and NextGenerationEU (European Union);
A*MIDEX, ANR, IPhU and Labex P2IO, and R\'{e}gion Auvergne-Rh\^{o}ne-Alpes (France);
AvH Foundation (Germany);
ICSC (Italy); 
GVA, XuntaGal, GENCAT, Inditex, InTalent and Prog.~Atracci\'on Talento, CM (Spain);
SRC (Sweden);
the Leverhulme Trust, the Royal Society
 and UKRI (United Kingdom).

\addcontentsline{toc}{section}{References}
\bibliographystyle{LHCb}
\bibliography{main,standard,LHCb-PAPER,LHCb-CONF,LHCb-DP,LHCb-TDR}

\newpage
\centerline
{\large\bf LHCb collaboration}
\begin
{flushleft}
\small
R.~Aaij$^{35}$\lhcborcid{0000-0003-0533-1952},
A.S.W.~Abdelmotteleb$^{54}$\lhcborcid{0000-0001-7905-0542},
C.~Abellan~Beteta$^{48}$,
F.~Abudin{\'e}n$^{54}$\lhcborcid{0000-0002-6737-3528},
T.~Ackernley$^{58}$\lhcborcid{0000-0002-5951-3498},
J. A. ~Adams$^{66}$\lhcborcid{0009-0003-9175-689X},
A. A. ~Adefisoye$^{66}$\lhcborcid{0000-0003-2448-1550},
B.~Adeva$^{44}$\lhcborcid{0000-0001-9756-3712},
M.~Adinolfi$^{52}$\lhcborcid{0000-0002-1326-1264},
P.~Adlarson$^{79}$\lhcborcid{0000-0001-6280-3851}
% ...

{\footnotesize \it

% ...
$^{35}$Nikhef National Institute for Subatomic Physics, Amsterdam, Netherlands\\
% ...
$^{44}$Instituto Galego de F{\'\i}sica de Altas Enerx{\'\i}as (IGFAE), Universidade de Santiago de Compostela, Santiago de Compostela, Spain\\
% ...
$^{48}$Physik-Institut, Universit{\"a}t Z{\"u}rich, Z{\"u}rich, Switzerland\\
% ...
$^{52}$H.H. Wills Physics Laboratory, University of Bristol, Bristol, United Kingdom\\
% ...
$^{54}$Department of Physics, University of Warwick, Coventry, United Kingdom\\
% ...
$^{57}$School of Physics and Astronomy, University of Glasgow, Glasgow, United Kingdom\\
$^{58}$Oliver Lodge Laboratory, University of Liverpool, Liverpool, United Kingdom\\
% ...
$^{66}$Syracuse University, Syracuse, NY, United States\\
% ...
$^{79}$Department of Physics and Astronomy, Uppsala University, Uppsala, Sweden, associated to $^{57}$\\
% ...
}
\end{flushleft}

\end{document}